\documentstyle[12pt]{article}

  \def\pp{{\mathchoice
          {%displaystyle
              \kern 1pt%
              \raise 1pt
              \vbox{\hrule width5pt height0.4pt depth0pt
                    \kern -2pt
                    \hbox{\kern 2.3pt
                          \vrule width0.4pt height6pt depth0pt
                          }
                    \kern -2pt
                    \hrule width5pt height0.4pt depth0pt}%
                    \kern 1pt
           }
            {%textstyle
              \kern 1pt%
              \raise 1pt
              \vbox{\hrule width4.3pt height0.4pt depth0pt
                    \kern -1.8pt
                    \hbox{\kern 1.95pt
                          \vrule width0.4pt height5.4pt depth0pt
                          }
                    \kern -1.8pt
                    \hrule width4.3pt height0.4pt depth0pt}%
                    \kern 1pt
            }
            {%scriptstyle
              \kern 0.5pt%
              \raise 1pt
              \vbox{\hrule width4.0pt height0.3pt depth0pt
                    \kern -1.9pt  %[e]=0.15pt
                    \hbox{\kern 1.85pt
                          \vrule width0.3pt height5.7pt depth0pt
                          }
                    \kern -1.9pt
                    \hrule width4.0pt height0.3pt depth0pt}%
                    \kern 0.5pt
            }
            {%scriptscriptstyle
              \kern 0.5pt%
              \raise 1pt
              \vbox{\hrule width3.6pt height0.3pt depth0pt
                    \kern -1.5pt
                    \hbox{\kern 1.65pt
                          \vrule width0.3pt height4.5pt depth0pt
                          }
                    \kern -1.5pt
                    \hrule width3.6pt height0.3pt depth0pt}%
                    \kern 0.5pt%}
            }
        }}

  \def\mm{{\mathchoice
                       {%displaystyle
                             \kern 1pt
               \raise 1pt    \vbox{\hrule width5pt height0.4pt depth0pt
                                  \kern 2pt
                                  \hrule width5pt height0.4pt depth0pt}
                             \kern 1pt}
                       {%textstyle
                            \kern 1pt
               \raise 1pt \vbox{\hrule width4.3pt height0.4pt depth0pt
                                  \kern 1.8pt
                                  \hrule width4.3pt height0.4pt depth0pt}
                             \kern 1pt}
                       {%scriptstyle
                            \kern 0.5pt
               \raise 1pt
                            \vbox{\hrule width4.0pt height0.3pt depth0pt
                                  \kern 1.9pt
                                  \hrule width4.0pt height0.3pt depth0pt}
                            \kern 1pt}
                       {%scriptscriptstyle
                           \kern 0.5pt
             \raise 1pt  \vbox{\hrule width3.6pt height0.3pt depth0pt
                                  \kern 1.5pt
                                  \hrule width3.6pt height0.3pt depth0pt}
                           \kern 0.5pt}
                       }}

\catcode`@=11
\def\un#1{\relax\ifmmode\@@underline#1\else
        $\@@underline{\hbox{#1}}$\relax\fi}
\catcode`@=12

\let\du=\du                     % dot-under
\def\d{\delta}
\def\e{\epsilon}

\def\k{\kappa}
\def\l{\lambda}
\def\m{\mu}
\def\n{\nu}
\def\o{\omega}
\def\p{\pi}

\def\r{\rho}
\def\s{\sigma}

\def\x{\xi}

\def\F{\Phi}
\def\G{\Gamma}

\def\O{\Omega}
\def\P{\Pi}

\def\ve{\varepsilon}

\def\cm{{\cal M}}

\def\car{{\cal R}}

\def\bo{{\raise-.15ex\hbox{\large$\Box$}}}               % D'Alembertian
\def\pa{\partial}                                       % curly d
\def\de{\nabla}                                         % del
\def\TH{{\raise.2ex\hbox{$\displaystyle \bigodot$}\mskip-4.7mu \llap H \;}}
\def\face{{\raise.2ex\hbox{$\displaystyle \bigodot$}\mskip-2.2mu \llap {$\ddot
        \smile$}}}                                      % happy face
\def\sp#1{{}^{#1}}                              % superscript (unaligned)
\def\leftrightarrowfill{$\mathsurround=0pt \mathord\leftarrow \mkern-6mu
        \cleaders\hbox{$\mkern-2mu \mathord- \mkern-2mu$}\hfill
        \mkern-6mu \mathord\rightarrow$}
\def\dvec#1{\vbox{\ialign{##\crcr
        \leftrightarrowfill\crcr\noalign{\kern-1pt\nointerlineskip}
        $\hfil\displaystyle{#1}\hfil$\crcr}}}           % <--> accent
\def\frac#1#2{{\textstyle{#1\over\vphantom2\smash{\raise.20ex
        \hbox{$\scriptstyle{#2}$}}}}}                   % fraction
\def\sfrac#1#2{{\vphantom1\smash{\lower.5ex\hbox{\small$#1$}}\over
        \vphantom1\smash{\raise.4ex\hbox{\small$#2$}}}} % alternate fraction
\def\bfrac#1#2{{\vphantom1\smash{\lower.5ex\hbox{$#1$}}\over
        \vphantom1\smash{\raise.3ex\hbox{$#2$}}}}       % "
\def\afrac#1#2{{\vphantom1\smash{\lower.5ex\hbox{$#1$}}\over#2}}    % "
\def\[{\lfloor{\hskip 0.35pt}\!\!\!\lceil}
\def\]{\rfloor{\hskip 0.35pt}\!\!\!\rceil}

\def\du#1#2{_{#1}{}^{#2}}
\def\ud#1#2{^{#1}{}_{#2}}

\def\fracm#1#2{\hbox{\large{${\frac{{#1}}{{#2}}}$}}}
\def\ha{{\fracmm12}}
\def\tr{{\rm tr}}
\def\Tr{{\rm Tr}}

\def\un{\underline}
\def\fracmm#1#2{{{#1}\over{#2}}}

\def\low#1{{\raise -3pt\hbox{${\hskip 0.75pt}\!_{#1}$}}}

\newskip\humongous \humongous=0pt plus 1000pt minus 1000pt
\def\caja{\mathsurround=0pt}
\def\eqalign#1{\,\vcenter{\openup2\jot \caja
        \ialign{\strut \hfil$\displaystyle{##}$&$
        \displaystyle{{}##}$\hfil\crcr#1\crcr}}\,}
\newif\ifdtup

\def\ref#1{$\sp{#1)}$}

\def\pl#1#2#3{Phys.~Lett.~{\bf {#1}B} (19{#2}) #3}
\def\np#1#2#3{Nucl.~Phys.~{\bf B{#1}} (19{#2}) #3}

\def\cmp#1#2#3{Commun.~Math.~Phys.~{\bf {#1}} (19{#2}) #3}

\def\ibid#1#2#3{{\it ibid.}~{\bf {#1}} (19{#2}) #3}

\parskip=\medskipamount                 % space between paragraphs (LaTeX)
\thicklines                         % thick straight lines for pictures (LaTeX)

\begin{document}

% =========================== UH title page ==========================

\thispagestyle{empty}               % no heading or foot on title page (LaTeX)

\def\border{                                            % UH border
        \setlength{\unitlength}{1mm}
        \newcount\xco
        \newcount\yco
        \xco=-24
        \yco=12
        \begin{picture}(140,0)
        \put(-20,11){\tiny Institut f\"ur Theoretische Physik Universit\"at
Hannover~~ Institut f\"ur Theoretische Physik Universit\"at Hannover~~
Institut f\"ur Theoretische Physik Hannover}
        \put(-20,-241.5){\tiny Institut f\"ur Theoretische Physik Universit\"at
Hannover~~ Institut f\"ur Theoretische Physik Universit\"at Hannover~~
Institut f\"ur Theoretische Physik Hannover}
        \end{picture}
        \par\vskip-8mm}

\def\headpic{                                           % UH heading
        \indent
        \setlength{\unitlength}{.8mm}
        \thinlines
        \par
        \begin{picture}(29,16)
        \put(75,16){\line(1,0){4}}
        \put(80,16){\line(1,0){4}}
      \put(85,16){\line(1,0){4}}
        \put(92,16){\line(1,0){4}}

        \put(85,0){\line(1,0){4}}
        \put(89,8){\line(1,0){3}}
        \put(92,0){\line(1,0){4}}

        \put(85,0){\line(0,1){16}}
        \put(96,0){\line(0,1){16}}
        \put(92,16){\line(1,0){4}}

        \put(85,0){\line(1,0){4}}
        \put(89,8){\line(1,0){3}}
        \put(92,0){\line(1,0){4}}

        \put(85,0){\line(0,1){16}}
        \put(96,0){\line(0,1){16}}
        \put(79,0){\line(0,1){16}}
        \put(80,0){\line(0,1){16}}
        \put(89,0){\line(0,1){16}}
        \put(92,0){\line(0,1){16}}
        \put(79,16){\oval(8,32)[bl]}
        \put(80,16){\oval(8,32)[br]}

        \end{picture}
        \par\vskip-6.5mm
        \thicklines}

\border\headpic {\hbox to\hsize{
\vbox{\noindent DESY ~96 -- 071 \hfill April 1996    \\
ITP--UH -- 05/96      \hfill            hep-th/9604141 \\
ISSN 0418 --9833     \hfill            revised version       }}}
\noindent
\vskip1.3cm
\begin{center}

{\Large\bf One--Loop~ Finiteness~ of~ the~ Four-Dimensional  
\vglue.1in                                                                  
           Donaldson-Nair-Schiff~ Non-Linear~ Sigma-Model~\footnote{Supported 
by the `Volkswagen Stiftung'}}\\
\vglue.3in

Sergei V. Ketov \footnote{
On leave of absence from:
High Current Electronics Institute of the Russian Academy of Sciences,
\newline ${~~~~~}$ Siberian Branch, Akademichesky~4, Tomsk 634055, Russia}

{\it Institut f\"ur Theoretische Physik, Universit\"at Hannover}\\
{\it Appelstra\ss{}e 2, 30167 Hannover, Germany}\\
{\sl ketov@itp.uni-hannover.de}
\end{center}
\vglue.2in
\begin{center}
{\Large\bf Abstract}
\end{center}

The most general four-dimensional non-linear sigma-model, having the 
second-order derivatives only and interacting with a background metric and 
an antisymmetric tensor field, is constructed. Despite its apparent 
non-renormalizability, just imposing the one-loop UV-finiteness conditions 
determines the unique model, which may be finite to all orders of 
the quantum perturbation theory. This model is known as the four-dimensional 
Donaldson-Nair-Schiff theory, which is a four-dimensional analogue of the 
standard two-dimensional Wess-Zumino-Novikov-Witten model, and whose unique 
finiteness properties and an infinite-dimensional current algebra have long 
been suspected.

\newpage
\hfuzz=10pt

{\bf 1} {\it Introduction}. The two-dimensional Wess-Zumino-Novikov-Witten 
(WZNW) model \cite{wznw} is the particular non-linear sigma-model (NLSM) whose
target space is a group manifold, and the NLSM torsion to be represented by 
the WZ term parallelizes the group manifold. The WZNW model is a conformally 
invariant quantum field theory and, hence, it is finite to all orders of the 
quantum perturbation theory. It possesses on-shell the conserved affine 
currents which satisfy an infinite-dimensional affine algebra.~\footnote{See, 
e.g. ref.~\cite{kbook} for a review.}

It is quite natural to investigate to what extent those nice properties can be 
generalized to four dimensions, which would allow one to generalize some 
familiar concepts of two-dimensional conformal field theory up to four 
dimensions. Recently, some progress along these lines was reported by Losev, 
Moore, Nekrasov and Shatashvili~\cite{lmns}. They mostly discussed the 
algebraic geometry aspects of a possible four-dimensional generalization of 
the WZNW model, while the issues of its renormalization and anticipated 
UV-finiteness remained open. In this Letter, I investigate the one-loop 
renormalization of the general four-dimensional NLSM coupled to a 
four-dimensional metric and a two-form, and determine the unique class of 
models which are one-loop finite. The relevant NLSM is essentially the one 
first introduced by Donaldson~\cite{d}, and later studied by Nair and 
Schiff~\cite{ns} in the context of the three-dimensional K\"ahler-Chern-Simons
 theory. 
\vglue.2in

{\bf 2} {\it The general action, and the background field method}. Let $R_4$ 
be a four-dimensional manifold of Euclidean signature, which is parameterized 
by the coordinates $x^{\m}$, $\m=1,2,3,4$, and is equipped with a metric 
$h_{\m\n}(x)$ {\it and} a 2-form $\o=\o_{\m\n}(x) dx^{\m}\wedge dx^{\n}$, 
$\o_{\m\n}=-\o_{\n\m}$. Let $\F$ be a map from $R_4$ to another 
$n$-dimensional manifold $\cm$, and $\F^a$, $a=1,2,\ldots,n$, be the 
coordinates on $\cm$.

The most general NLSM action which is invariant under the reparametrizations
of both $R_4$ and $\cm$, and has only second-order derivatives of the fields 
$\F^a$, is given by
$$ I[\F;h,\o] = \fracmm{1}{2\l^2} \int d^4x\,\sqrt{h}h^{\m\n}\pa_{\m}
\F^a\pa_{\n}
\F^b g_{ab}(\F) + \fracmm{\k}{2\l^2}\int d^4 x\,\ve^{\m\n\l\r}\o_{\m\n}\pa_{\l}
\F^a \pa_{\r}\F^b b_{ab}(\F)~,\eqno(1)$$
where the NLSM target space metric $g_{ab}$ and a 2-form $b=b_{ab}(\F) d\F^a
\wedge d\F^b$, $b_{ab}=-b_{ba}$, on $\cm$ have been introduced. 
In our notation, $dx^{\m}$ carries dimension one. Accordingly, the NLSM 
coupling constant $\l$ is of dimension one too, whereas another coupling 
constant $\k$ and all the fields are dimensionless.

In order to make our quantum calculations covariant with respect to the target 
space metric, we use the covariant background field method suitable for the 
NLSM \cite{he,k2}. Let $\r^a(x,s)$ be the geodesic connecting 
$\F^a(x)$ with $\F^a(x) 
+\p^a(x)$, such that $\r^a(x,s=0)=\F^a(x)$, $\r^a(x,s=1)=\F^a(x)+\p^a(x)$, and
$$ \fracmm{d^2}{ds^2}\r^a + \G^a_{bc}[\r]\fracmm{d}{ds}\r^b\fracmm{d}{ds}\r^c
=0~, \eqno(2)$$ 
where $\G^a_{bc}$ are the Christoffel symbols with respect to the metric 
$g_{ab}$. Let $\vec{\x}$ be the tangent vector to the geodesic, i.e.
$$ \x^a_{\rm s} =\fracmm{d}{ds}\r^a~,\qquad{\rm and}\qquad
  \left. \x^a_{\rm s}\right|_{s=0}= \x^a~.\eqno(3)$$
The fields $\x^a(x)$ will be considered as the fundamental quantum fields
in our theory whereas $\F^a$, $h_{\m\n}$ and $\o_{\m\n}$ as the background 
fields.

Let us now expand $I[\F+\p(\x)]$ in terms of the $\x$-fields, which ensures 
that all the coefficients of the expansion are tensorial quantities on $\cm$,
$$ I [ \F+\p(\x )]= I[ \F ] + I_1 + I_2 + \ldots~,\quad {\rm where}\quad
I_n=\left. \fracmm{1}{n!}\fracmm{d^n}{ds^n}I[\r(s)]\right|_{s=0}~.\eqno(4)$$
In practice, it is more convenient to use the derivative $D(s)$ to be defined
with the covariant completion, instead of $d/ds$, i.e.
$$\eqalign{
D(s) S[\r]=~&~\fracmm{d}{ds} S[\r]~,\cr
D(s)W_a[\r]=~&~\fracmm{d}{ds} W_a[\r] - \G^c_{ab}\x^b_{\rm s}W_c[\r]~,\cr
D(s)W^a[\r]=~&~\fracmm{d}{ds}W^a[\r] + \G^a_{bc}\x^b_{\rm s}W^c[\r],\cr}
\eqno(5)$$
Obviously, if $D(s)$ acts on a tensor function of $\r(s)$ only, we have 
$D(s)=\x^a_{\rm s}D_a$. Here are some useful identities:
$$\fracmm{d}{ds}\pa_{\m}\r^a=\pa_{\m}\x^a_{\rm s}~,\quad
\fracmm{d}{ds}g_{ab}=\x^c_{\rm s}\pa_cg_{ab}~,\quad D(s)g_{ab}=0~,\quad
D(s)\x^a_{\rm s}=0~,\eqno(6)$$
and
$$ D(s)\pa_{\m}\r^a=(D_{\m}\x_{\rm s})^a~,\quad D^2(s)\pa_{\m}\r^a=R\ud{a}{bcd}
[\r]\x^b_{\rm s}\x^c_{\rm s}\pa_{\m}\r^d~,\eqno(7)$$
where
$$ (D_{\m}\x_{\rm s})^a \equiv
\pa_{\m}\x^a_{\rm s} +\G^a_{bc}[\r]\pa_{\m}\F^b\x^c_{\rm s}~.
\eqno(8)$$

The first term in the expansion (4) (and, hence, all the other) can only be 
expressed in terms of the totally antisymmetric field strength (3-form) 
$H_{abc}\equiv\frac{3}{2}\pa_{[a}b_{bc]}$ of the potential (2-form) 
 $b_{ab}$ if 
we require the 2-form $\o$ to be {\it closed}, after integrating by parts 
in the second term of eq.~(1). A calculation yields
$$ I_1=\fracmm{1}{\l^2}\int d^4 x\,\sqrt{h}h^{\m\n}(D_{\m}\x)^a\pa_{\n}
\F^bg_{ab}
(\F) +\fracmm{\k}{\l^2}\int d^4x\,\ve^{\m\n\r\l}\o_{\m\n}\pa_{\r}\F^a\pa_{\l}
\F^b\x^c H_{abc}(\F)~.\eqno(9)$$

Similarly, we find
$$\eqalign{
I_2 =~&~ \fracmm{1}{2\l^2} \int d^4x\,\sqrt{h}h^{\m\n}\left[ g_{ab}(D_{\m}\x)^a
(D_{\n}\x)^b + R_{abcd}\pa_{\m}\F^a\pa_{\n}\F^d\x^b\x^c\right] \cr
~&~+ \fracmm{\k}{2\l^2}\int d^4x\,\ve^{\m\n\l\r}\o_{\m\n}\left[ 
2H_{abc}(\F)\pa_{\l}\F^a(D_{\r}\x)^b\x^c +D_bH_{adc}\pa_{\l}
\F^a\pa_{\r}\F^d\x^b\x^c\right]~.\cr} \eqno(10)$$

Let $V^i_a(\F)$ be the {\it vielbein} associated with the NLSM target space 
metric $g_{ab}(\F)$,
$$ g_{ab}(\F)=V^i_a(\F)V^j_b(\F)\d_{ij}~,\qquad V^{ai}=g^{ab}V^i_b~.\eqno(11)$$
Then, we can define the new derivative $\de_{\m}$ (without torsion) as
$$ V^i_a(D_{\m}\x)^a \equiv (\de_{\m}\x)^i = (\d^{ij}\pa_{\m}
+A^{ij}_{\m})\x^j~,\eqno(12)$$
where we have also introduced the vector $\x^i$ associated with $\x^a$ 
as follows:
$$ \x^i=V^i_a\x^a~,\qquad \x^a=V^{ai}\x^i~.\eqno(13)$$
It now allows us to rewrite eq.~(10) to the form
$$\eqalign{
I_2 =~&~\fracmm{1}{2\l^2}\int d^4x\,\left\{ \sqrt{h}h^{\m\n}(\de_{\m}\x)^i
(\de_{\n}\x)^i +2\k\ve^{\m\n\l\r}\o_{\m\n}H_{aij}\pa_{\l}\F^a(\de_{\r}\x)^i\x^j
\right.\cr
~&~\left. + (\sqrt{h}h^{\m\n}R_{aijb}\pa_{\m}\F^a\pa_{\n}\F^b
+\k\ve^{\m\n\l\r}\o_{\m\n}D_iH_{abj}\pa_{\l}\F^a\pa_{\r}\F^b)\x^i\x^j\right\}~,
\cr }\eqno(14)$$
where the kinetic terms for the quantum fields $\x^i$ have the standard form, 
thus defining the usual ($R_4$-covariant) $1/p^2$ propagator.

We are now going to redefine the connection in $\de_{\m}$ to 
$\hat{\de}_{\m}$, in order to hide in it the $(\de\x)\x$ term appearing in 
eq.~(14). It suffices to take
$$ \hat{\de}_{\m}\equiv \de_{\m} +X_{\m}~,\eqno(15)$$
where
$$ X_{\m}^{ij} = \fracmm{\k}{\sqrt{h}}\ve^{\r\l\s\n}h_{\m\n}\o_{\r\l}\pa_{\s}
\F^a H^{ij}_a~.\eqno(16)$$

Finally, we have
$$\eqalign{
I_2=~&~\fracmm{1}{2\l^2}\int d^4x\,\sqrt{h} \left\{ 
h^{\m\n}(\hat{\de}_{\m}\x)^i(\hat{\de}_{\n}\x)^i \right.\cr
~&~\left. + ( h^{\m\n}R_{aijb} +2\k\tilde{\o}^{\m\n}D_iH_{abj} 
+4\k^2H_{aki}H_{bkj}
\tilde{\o}^{\m\r}h_{\r\l}\tilde{\o}^{\l\n})\pa_{\m}\F^a\pa_{\n}\F^b\x^i\x^j
\right\}~,\cr}\eqno(17)$$
where we have introduced the dual $\tilde{\o}$ of $\o$ as
$$\tilde{\o}^{\m\n}\equiv \fracmm{1}{2\sqrt{h}}\ve^{\m\n\l\r}\o_{\l\r}~.
\eqno(18)$$

It is straightforward to calculate the other terms in the background-quantum 
field expansion (4) at any given order, although no simple reccursion formula 
is known. The $n$-th order term has the structure
$$\eqalign{
I_n=~&~\fracmm{1}{2\l^2}\int d^4x\,\sqrt{h}\left\{ 
\left( h^{\m\n}\P^{(n,2)}_{(i_1\cdots i_n)(ab)} 
+ \tilde{\o}^{\m\n}E^{(n,2)}_{(i_1\cdots i_n)[ab]}\right)
\x^{i_1}\x^{i_2}\cdots\x^{i_n}\pa_{\m}\F^a\pa_{\n}\F^b \right. \cr 
~&~ + \left(h^{\m\n}\P^{(n,1)}_{(i_1\cdots i_{n-1})ja} +\tilde{\o}^{\m\n}
E^{(n,1)}_{(i_1\cdots i_{n-1})ja}\right) \x^{i_1}\x^{i_2}\cdots\x^{i_{n-1}}
(\hat{\de}_{\m}\x)^j\pa_{\n}\F^a \cr 
~&~ \left. +\left(h^{\m\n}\P^{(n,0)}_{(i_1\cdots i_{n-2})(j_1j_2)}
+ \tilde{\o}^{\m\n}E^{(n,0)}_{(i_1\cdots i_{n-2})[j_1j_2]}\right)
\x^{i_1}\x^{i_2}\cdots\x^{i_{n-2}}
(\hat{\de}_{\m}\x)^{j_1}(\hat{\de}_{\n}\x)^{j_2} 
\right\}~, \cr}\eqno(19)$$
where the $\P$'s and $E$'s are certain tensors to be constructed in terms of 
the curvature $R_{ijkl}$, the torsion $H_{ijk}$ and the covariant derivatives 
$\hat{D}_i$ on $\cm$, the metric $h$ and the two-form $\tilde{\o}$ on 
$R_4$.~\footnote{In general, it is {\it not} possible to express the full 
expansion solely in terms of the generalized (with \newline ${~~~~~}$
torsion) curvature $\hat{R}$ and the generalized (with torsion) covariant 
derivative $\hat{D}_i$, i.e. without an 
\newline ${~~~~~}$ explicit appearance of the torsion $H$.}
\vglue.2in

{\bf 3} {\it One-loop finiteness conditions and their solution}. We are now in
a position to investigate the NLSM {\it one-loop} finiteness conditions, 
since eq.~(17) is already sufficient for that purpose. We temporarily put aside
the issues of the $R_4$-background dependence of the quantum effective action 
and the wave-function renormalization, i.e. consider on-shell contributions to
 the NLSM beta-functions first.

As far as the one-loop finiteness is concerned, the $(\pa\F)^2\x^2$ 
contribution to the  $I_2$ has to vanish since, otherwise, it would inevitably
 contribute to the beta-functions. In the absence of the second term in the 
action (1), it would lead to the vanishing NLSM curvature $R_{ijkl}$ and, 
hence, to the linear NLSM action. The well-known statement about the 
non-renormalizabilty of the standard four-dimensional NLSM is therefore 
recovered this way. In the presence of the second term in the action (1), the
situation is different since we now have the two additional resources --- 
the 2-form $\o$ on $R_4$ and the torsion 3-form $H$ on $\cm$ --- and they both 
can be adjusted in such a way that a cancellation between the first and the 
third terms in front of the $(\pa\F)^2\x^2$ contribution becomes possible. 
Clearly, it can only happen if we constrain the two-form $\o$ 
to satisfy the equation
$$  h^{\m\n} = -\tilde{\o}^{\m\r}h_{\r\l}\tilde{\o}^{\l\n}~.\eqno(20)$$
Given eq.~(20), we still need the relation
$$ R_{ijkl} + 4\k^2 H_{ijm}H_{lkm}=0~.\eqno(21)$$
The minus sign in eq.~(20) is important since, otherwise, there is no solution
(see eq.~(26) below).

The second term in front of the  $(\pa\F)^2\x^2$ in eq.~(17) has to vanish
separately, since it represents an independent structure. It adds another 
condition,
$$ D_iH_{jkl}=0~,\eqno(22)$$
i.e. the torsion on $\cm$ has to be covariantly constant.

The remaining contribution to the $I_2$ now amounts to the {\it minimal} 
coupling of the quantum $\x$-fields to the external composite gauge field 
$X_{\m}$. Because of the gauge invariance, the on-shell invariant counterterms
are to be constructed in terms of the generalized field strength and its 
covariant derivatives. In the case under consideration, the one-loop 
finiteness requires a {\it parallelizable} manifold $\cm$. The 
parallelizability condition for the  generalized curvature $\hat{R}_{ijkl}$ is
also sufficient since it is equivalent to eqs.~(21) and (22) at the certain 
value of the real parameter $\k$ (see below). 

Since the only parallelizable manifolds are group manifolds and seven-sphere, 
we have to choose our torsion to be represented by the group structure 
constants,~\footnote{The case of the seven-sphere deserves a separate study, 
and it is not considered here.}
$$H_{ijk}=f_{ijk}~,\eqno(23)$$
which automatically fulfils the condition (22). Indeed, for a group manifold, 
the vielbein is known to satisfy the Maurer-Cartan equation
$$ \fracmm{\pa V^i_a}{\pa\F^b} -  \fracmm{\pa V^i_b}{\pa\F^a}
+2f^{ijk}V^j_aV^k_b=0~,\eqno(24)$$
whereas both the torsion and the curvature components, $H_{ijk}$ and 
$R_{ijkl}$, are all constants. Taking into account the explicit formula for 
the group curvature in terms of the structure constants,~\footnote{
We assume that our group is semi-simple and compact, for simplicity.}
$$ R_{ijkl}=-f^m_{ij}f^m_{kl}~,\eqno(25)$$
eq.~(21) amounts to the relation
$$ 4\k^2=1,\quad {\rm or} \quad \k=\pm \fracm{1}{2}~,\eqno(26)$$
which determines the dimensionless parameter $\k$ up to a sign. The 
dimensionful coupling constant $\l$ remains arbitrary, and it does not play an
 essential role in our considerations. 

The crucial equation (20) has very simple geometrical meaning. Let us define
the new tensor
$$ \tilde{\o}^{\m\r}h_{\r\l}\equiv J\ud{\m}{\l}~.\eqno(27)$$
Eq.~(20) can now be rewritten to the form
$$ J\ud{\m}{\l}J\ud{\l}{\r}= -\d\ud{\m}{\r}~,\eqno(28)$$
i.e. $J$ is nothing but an (almost) {\it complex structure} on $R_4$~! 
Eq.~(27) also
implies that $J^{\m\n}$ is antisymmetric, which means that $R_4$ has to be a
{\it hermitian} manifold. Note that the one-loop finiteness conditions do 
{\it not} imply $\de_{\m}J\ud{\l}{\r}=0$ or $\hat{\de}_{\m}J\ud{\l}{\r}=0$, so 
that $R_4$ does not necessarily need to be a K\"ahler manifold. If, however, 
$R_4$ is K\"ahlerian, then the closure of $\o$ follows automatically.

To summarize, the one-loop on-shell finiteness conditions for the 
four-dimensional NLSM of eq.~(1) are:\\
(i) the four-dimensional `spacetime' $R_4$ has to be a {\it hermitian} 
manifold, equipped with a {\it closed} two-form $\o_{\m\n}$ defined by 
eq.~(24), i.e. {\it dual} to $J^{\m\n}$; \\
(ii) the NLSM target space manifold $\cm$ has to be a parallelizable 
{\it group} manifold (or, maybe, a seven-sphere).

The alternative derivation of the one-loop counterterm for the theory (17) can
 be performed by the (generalized) Schwinger-de Witt method \cite{bav}. The 
relevant action (17) can be represented in the form
$$ I_2 = \fracmm{1}{2\l^2}\int d^4x\,\sqrt{h}\,
\x^i\hat{F}_{ij}(\hat{\de})\x^j~,\eqno(29)$$
where the minimal second-order differential operator $\hat{F}(\hat{\de})$ can 
be easily read off from eq.~(17). It is now straightforward to extract the 
one-loop counterterm for our case from the general results of ref.~\cite{bav}.
Having used dimensional regularization with the regularization parameter 
$2\e=4-d$, we find
$$\eqalign{
\left.-\ha\Tr\ln \hat{F}(\hat{\de})\right|_{\rm div.}=~&~
\fracmm{i}{32\p^2\e}\int d^4x\,\sqrt{h}\,
\tr\left\{\fracm{1}{180}\left(R_{\m\n\l\r}
R^{\m\n\l\r}-R_{\m\n}R^{\m\n}+\bo R\right)\hat{\bf 1}\right.\cr
~&~ \left.+\fracm{1}{2}\hat{P}^2 
+\fracm{1}{12}\hat{\car}_{\m\n}\hat{\car}^{\m\n}
+\fracm{1}{6} \bo \hat{P}\right\}~,\cr}\eqno(30)$$
where $\hat{P}$ just represents the  $(\pa\F)^2$ term in eq.~(17), whereas the
generalized gauge field-strength $\hat{\car}^2$ is proportional to the 
generalized curvature with torsion. 

Eq.~(30) actually tells us something more, namely, the dependence of the 
one-loop
counterterm from the four-dimensional background `spacetime' metric $g_{\m\n}$ 
also. Because of the Gauss-Bonnet theorem valid in four dimensions, the
curvature-squared term is reducible to the Ricci-tensor-dependent terms. Hence,
in order to cancel all the $R_4$ background curvature dependent terms in the 
one-loop counterterm, it is necessary and sufficient to impose the 
{\it Ricci-flateness} condition on the $R_4\,$.

As regards possible (non-linear) field-renormalization effects, they should 
{\it not} be relevant for our results. As far as a NLSM is concerned, there 
exists a quantum field parametrization in which the field renormalization is 
absent (since we were not imposing any restrictions on allowed quantum field
parametrizations, the parametrization required is just the one determineed by 
actual renormalization) \cite{k2}.

For any complex manifold $R_4$ one can choose complex coordinates 
$(z^i,z^{\bar{i}})$, where $z^{\bar{i}}=(z^i)^*$ and $i=1,2$, in such a way 
that the complex structure $J$ takes the canonical constant form. Given such 
complex coordinates, the action (1) with $\cm=G$ takes the form of the
Donaldson-Nair-Schiff (DNS) action~\cite{d,ns}
$$ I_{\rm DNS}[g] =-\,\fracmm{i}{4\p}\int_{R_4}\,\o\wedge\tr(g^{-1}\pa g 
\wedge g^{-1}\bar{\pa}g) +\fracmm{i}{12\p}\int_{R_5}\,
\o\wedge\tr(g^{-1}dg)^3~,\eqno(31)$$
where we have introduced $R_5=R_4\otimes[0,1]$ and the $G$-valued 
fields~\footnote{The generators $t^i$ of the Lie algebra of $G$ with
 the structure constants $f_{ijk}$ satisfy the relations 
\newline ${~~~~~}$ $\[t^i,t^j\]=2if^{ijk}t^k$ and $\tr(t^it^j)=2\d^{ij}$.}
$$ g(x)=\exp\left[i\F^i(x)t^i\right]~.\eqno(32)$$

In accordance with the one-loop finiteness conditions, the 2-form
$$ \o=\fracmm{i}{2\l^2}h_{i\bar{j}}dz^i\wedge dz^{\bar{j}}\eqno(33)$$
has to be closed. If $R_4$ is a K\"ahler manifold, there exists the natural 
2-form $\o$ which satisfies all our conditions -- the so-called {\it K\"ahler} 
form~\cite{cw}.
\vglue.2in

{\bf 4} {\it All-loop finiteness of the DNS action~?} To prove the all-loop 
(on-shell, or $S$-matrix) finiteness of the DNS action, let us
return back to the background field expansion specified by eqs.~(4) and (19). 
Under the conditions (i) and (ii) given in sect.~3, only the last terms in the
third line of eq.~(19) survive, i.e.
$$ \P^{(n,2)}=\P^{(n,1)}=E^{(n,2)}=E^{(n,1)}=0~,\eqno(34)$$
the non-vanishing tensors $\P^{(n,0)}$ and $E^{(n,0)}$ being the products of 
the group structure constants~\cite{k2}. Eq.~(34) means, in particular, that 
{\it all} the 
$\pa\F$ dependence in the background-quantum field expansion (4) can be hidden 
inside the covariant derivatives (with torsion). Then, as far as the $l$-loop 
counterterms are concerned, the 
$\pa\F$-dependent contributions can only show up via the generalized field 
strength which is vanishing in our case. Hence, no covariant counterterms 
actually appear modulo the ones coming from the vacuum diagrams and depending 
upon the `spacetime' metric only. The rigorous finiteness proof would require
taking into account possible non-covariant divergences, if any. I believe that
they all can be removed by a wave-function renormalization. Finally, in order 
to make sure of the absence of one-loop vacuum divergences. we already know 
that our complex `spacetime' should be Ricci-flat. If it is also a K\"ahlerian
manifold, it is then a {\it hyper-K\"ahlerian} one. It is the well-known 
theorem in four dimensions that any hyper-K\"ahlerian manifold is 
actually {\it self-dual}~\cite{ah}. But, for the self-dual four-dimensional 
backgrounds there can be no vacuum counterterms at all~! It can be proved e.g.,
by using the gravitational background field expansion for the self-dual 
gravity~\cite{gr}, or by invoking the relation which exists between the 
self-dual gravity and critical N=2 strings~\cite{ov,kbook}. The self-dual 
gravity is the {\it exact} effective field theory of the closed N=2 strings in
four dimensions,~\footnote{Perhaps, modulo non-perturbative N=2 string 
corrections.} while all the N=2 string scattering amplitudes with more than 
three legs vanish~\cite{bv}. Therefore, there cannot be any renormalization of 
the self-dual gravity in four dimensions, at any loop order. 

Above, both the Ricci-flatness and the K\"ahlerness conditions for $R_4$ were 
explicitly used. It may well be possible to require only the  Ricci-flatness 
for the hermitian `spacetime', in order to get rid of the vacuum divergences, 
after switching to the (N=1) supersymmetric version of the theory. I believe
that the DNS action can be supersymmetrized up to $N=4$.

The connection of the DNS action to the theory of N=2 strings becomes explicit
when considering the equations of motion following from the action (31):
$$ \o\wedge \bar{\pa}(g^{-1}\pa g)=0~.\eqno(35)$$
These equations describe the coupling of the self-dual gravity to be 
represented by $\o$ to the principal NLSM fields associated with the self-dual
Yang-Mills theory. Indeed, the so-called {\it Yang} equations 
$\bar{\pa}(g^{-1}\pa g)=0$ following from  eq.~(35) are known to be equivalent
to the self-dual Yang-Mills equations in a particular gauge~\cite{yang}. Since
the self-dual Yang-Mills theory is the {\it exact} effective field theory for 
the open N=2 strings,~${}^8$ eq.~(35) can be interpreted as the exact 
effective equation describing the interaction of open and closed N=2 strings. 
Invoking now the vanishing theorems for the open {\it and} closed N=2 string 
amplitudes, an anticipated finiteness of the DNS theory does not seem to be 
very surprising.
\vglue.2in

{\bf 5} {\it Conclusion}. A possible finiteness of the DNS theory is 
consistent with 
its classical integrability (the Yang equations are solvable like that of the 
self-dual Yang-Mills~!). After an explicit `space-time' splitting of the
Euclidean manifold $R_4=R_3\otimes R_1$, and introducing the phase space 
$\{P^i(x),g(x)\}$ for the theory (31), where the momenta $P^i$ are defined 
with respect to the symplectic form~\cite{lmns}
$$ \O_{\o}=\int_{R_3}\,\tr\left[\d P\wedge g^{-1}\d g 
- (I+\fracm{1}{4\p}\o\wedge g^{-1}dg)(g^{-1}\d g)^2\right]~,\eqno(36)$$
one finds the commutation relations
$$\eqalign{
\[P^i(x),P^j(y)\]_{\o}=~&~f\du{k}{ij}(I+\fracm{1}{4\p}\o\wedge 
g^{-1}dg)^k\d^{(3)} (x-y)~,\cr
\[P^i(x),g(y)\]_{\o}=~&~g(y)t^i\d^{(3)}(x-y)~.\cr}\eqno(37)$$
The infinite-dimensional symmetry algebra of the DNS theory can now be 
elegantly expressed in terms of the charges
$$ Q(\ve)=\int_{R_3}\,\ve^i\left[ I^i-\fracm{1}{4\p}\o\wedge 
(g^{-1}dg)^i\right]~, \eqno(38)$$
where the Lie algebra-valued functional parameters $\ve^i(x)$ have been 
introduced. The charges $ Q(\ve)$ satisfy the algebra~\cite{lmns}
$$\{Q(\ve_1),Q(\ve_2)\}=Q(\[\ve_1,\ve_2\])+\int_{R_3}\,\o\wedge\tr
(\ve_1d\ve_2)~,\eqno(39)$$
which is the four-dimensional analogue of the affine algebra in the WZNW 
theory.

Given such remarkable properties of the DNS theory, there should exist its free
field representation to be obtained by a field redefinition of the fields 
$g(x)$, which is yet to be found. Also, I expect the DNS action to be 
connected to a theory of $2+2$ dimensional membranes, or the M-theory. Even if 
the DNS theory is not finite beyond one loop, its supersymmetric version may 
appear to be finite to all orders.
\vglue.2in

{\bf Acknowledgements}

I would like to thank the Theory Group of the Institute of Experimental and
Theoretical Physics (ITEP), where this work was initiated, for hospitality 
extended to me during my visit to Moscow, as well as to thank I. Buchbinder, 
A. Deriglazov, A. Losev and A. Morozov for useful discussions.
\vglue.2in

\end{document}